\documentclass[aps,twocolumn,pra,showpacs,floatfix]{revtex4}
\usepackage{epsfig}
\usepackage{graphicx}
\usepackage{dcolumn}
\usepackage{amssymb,amsmath}
\usepackage{mathrsfs}
\begin{document}

\title{Effects of electrons on nuclear clock transition frequency in  $^{229}$Th  ions}

\author{V. A. Dzuba and V. V. Flambaum}

\affiliation{School of Physics, University of New South Wales, Sydney 2052, Australia}

\begin{abstract}
We perform calculations of the energy shift of the nuclear clock transition frequency $^{229}$Th as a function of the number of electrons in Th ion.
  We demonstrate that the dependence of the nuclear frequency on electron configuration is significant. E.g., removing one electron from the atom leads to relative shift of the nuclear frequency $\sim 10^{-7}$, which is twelve orders of magnitude larger than expected relative uncertainty of the nuclear clock transition frequency ($\sim 10^{-19}$). This leads to difference of the nuclear clock frequencies in Th~IV, Th~III, Th~II and Th~I.
  The relative change of the nuclear frequency between neutral Th and its bare nucleus is 1\%. We also calculate the field shift constants for isotopic and isomeric shifts of atomic electron transitions in Th ions. 
\end{abstract}

\maketitle

Nucleus of the $^{229}$Th isotope has a unique feature of having very low-energy excitation connected to the ground state by the magnetic dipole (M1) transition (see, e.g. Reviews~\cite{Rev,Rev1} and references therein). The latest, most precise measurements, give the value of 8.338(24)~eV~\cite{wn} (see also \cite{wn1,wn2,wn3,wn4,wn5}) for the energy of this excitation, which is very small on nuclear scale. This feature attracted many researches for plans to build nuclear clock of exceptionally high accuracy  - see e.g. ~\cite{PeikTamm,Thm}.
The projected relative uncertainty is expected to reach $10^{-19}$~\cite{Th3+}. In addition, there are strong arguments that this nuclear clock would be very sensitive to physics beyond standard model including space-time variation of the fundamental constants, violation of the Lorentz invariance and Einstein equivalence principle, and search  for scalar and axion dark matter fields ~\cite{vara,Lorentz,varq,vara1,vara2,vara3,vara4,Arvanitaki,Stadnik}.
There are plans to use Th ions of different ionisation degree~\cite{Th3+,Th+,dr} and even solid-state Th nuclear clock~\cite{Hudson,ThSS,ThSS1}.
In this work we show that in all these systems the frequency of the nuclear clock will be different. This is due to the Coulomb interaction of atomic electrons with the nucleus, leading to the significant 
electronic shift of the nuclear transition frequency. There is also a smaller shift due to the magnetic interaction.

This electronic  shift depends on electron configuration and it is different in different systems, like Th~IV, Th~III, Th~II and Th~I , leading to different nuclear frequencies.
This shift for electronic state $a$ is given by
\begin{equation}\label{e:IS}
\Delta E_a = F_a \delta \langle r^2 \rangle,
\end{equation}
where $F_a$ is the field shift constant of state $a$ which can be obtained from atomic calculations; $\delta \langle r^2 \rangle$ is the change of the nuclear root-mean square radius between the excited and ground nuclear states. The most accurate value for $\delta \langle r^2 \rangle$ was recently derived in Ref.~\cite{dr}, 
$^{229m,229}\delta \langle r^2 \rangle=0.0105(13)~{\rm fm}^2$. This enables us to determine the electronic  shift of nuclear transition  frequency for different thorium systems by calculating the field shift constants $F_a$ and using (\ref{e:IS}). For example, difference of the nuclear frequencies between Th~III and Th~IV is given by 
\begin{equation}\label{e:ISN}
\Delta \omega_N = (F_a (\text{Th~III}) -  F_a(\text{Th~IV}) ) \delta \langle r^2 \rangle,
\end{equation}
State $a$ in this case is the ground electronic state of the ion.
 
 Note that  these field shift constants $F$ appear  also in the calculations of the isotopic and isomeric field shifts of electronic  transition frequencies. The difference is that in the isotopic and isomeric shifts we need difference of $F$ for final state $b$ and initial state $a$  of the electronic transition. The nuclear state does not change in this electronic  transition. For isotope shift it is usually the ground nuclear state. For isomeric shift it is isomeric (excited) state or  ground   state  of the same nucleus. The isotopic and isomeric field shifts  of the electronic transition frequency are given by 
 \begin{equation}\label{e:ISab}
\Delta \omega_{ab} = (F_b  -  F_a) \delta \langle r^2 \rangle,
\end{equation}
 Numerical values of $\Delta \omega_N$ and  $\Delta \omega_{ab}$ can be calculated using values of the constants $F$ for different electron states  in Th~IV, Th~III, Th~II and Th~I presented in the Table \ref{t:F} . Note that we do not include  a contribution of  core electrons which cancels out in the difference of the values of $F$. For the isomeric shifts one may  use  $^{229m,229}\delta \langle r^2 \rangle=0.0105(13)~{\rm fm}^2$ measured in Ref.~\cite{dr}.

We use the combination of the single-double coupled cluster and the configuration interaction methods (SD+CI,~\cite{SD+CI}) and 
random-phase approximation (RPA)  method to perform the calculations. The SD+CI method gives us the wave functions, while the RPA method gives an effective operator of the field shift.
Corresponding equations have a form (see e.g. ~\cite{TDHF})
\begin{equation}\label{e:RPA}
(\hat H^{\rm HF} - \epsilon_c)\delta \psi_c = - (\hat F + \delta V_{\rm core}),
\end{equation}
where $H^{\rm HF}$ is the relativistic Hartree-Fock operator for the atomic core, index $c$ numerates single-electron states in the core, $\psi_c$ and 
$\delta \psi_c$ are corresponding single-electron functions and corrections due to the field shift operator $\hat F$, and $\delta V_{\rm core}$ is the change of the self-consistent Hartree-Fock potential due to the change in all core functions. Solving Eqs. (\ref{e:RPA}) self-consistently allows to determine  $\delta V_{\rm core}$. Note that the core is the same for Th~IV, Th~III, Th~II and Th~I. Therefore the SD+CI and RPA equations need to be solved only once. Then the field shift constant is given by
\begin{equation}\label{e:F}
F_a = \langle a|\hat F + \delta V_{\rm core} |a\rangle.
\end{equation}
We use hat to distinguish between the field shift constant $F$ and the field shift operator $\hat F = \delta V_{\rm nuc}/\delta \langle r^2 \rangle$.
The wave function $|a\rangle$ in (\ref{e:F}) is the many-electron wave function for valence electrons found in the SD+CI calculations. 
It has one, two, three or four valence  electrons.

The results of the calculations are presented in Table \ref{t:F}. We present energy levels and field shift constants for the ground and some excited states of 
Th~IV, Th~III, Th~II, and Th~I. We have chosen low-energy excited states and also some other  states of Th~III and Th~I for which other calculations and experimental data on isotope shift are available~\cite{dr}. 
% The energy levels are compared with experiment~\cite{Th-exp} to give some idea about the accuracy of the calculations.
The values of the field shift constants are compared with earlier calculations  in Ref. ~\cite{dr}. 

The difference  of  the field shift constants between our calculations and  calculations in Ref. ~\cite{dr} is few per cent. This difference may be used as an accuracy estimate since the calculations have been done by different methods.
 The largest difference is for the ground state of Th~II, which is 10\%.  However,  our number leads to more consistent results for values of  $\delta \langle r^2 \rangle$ extracted from the isotope shift measurements in ions  Th~II and  Th~III.  Indeed, using our numbers, $F$=49.6~GHz/fm$^2$ for the ground state and $F$=-29.1~GHz/fm$^2$ for the state at $E$=17122~cm$^{-1}$, 
for extracting the difference in root-mean-square radii $\delta \langle r^2 \rangle^{232,229}$ from the isotope shift data~\cite{dr} leads to the
%better consistency between the result for Th~II (our 
value  $\delta \langle r^2 \rangle^{232,229}$ = 0.321(32)~fm$^2$ (we assume 10\% uncertainty for the values of $F$), which is closer to
the data extracted from four transitions in Th~III (0.315(32), 0.312(42), 0.338(44), 0.322(53), see Table~I in~\cite{dr}). 
When all five numbers are taken into account, four numbers for Th~III from Ref.~\cite{dr} and our number for Th~II, 0.321(32), the final result is $\delta \langle r^2 \rangle^{232,229}$ = 0.320(15)~fm$^2$ (the final value of \cite{dr} is $\delta \langle r^2 \rangle^{232,229}$ = 0.299(15)~fm$^2$). 
Our result is in better agreement with the latest most accurate literature value $\delta \langle r^2 \rangle^{232,229}$=0.334(8)~fm$^2$ presented in Ref.~\cite{Marinova}. The new value of $\delta \langle r^2 \rangle^{232,229}$ leads to slightly different value of $\delta \langle r^2 \rangle^{229m,229}$.
Using the ratio of the isomeric and isotopic shifts from Ref.~\cite{Thm} we get $\delta \langle r^2 \rangle^{229m,229}$ = 0.0112(13)~fm$^2$. It is 7\% larger but agrees within error bars wth the value  $\delta \langle r^2 \rangle^{229m,229}$ = 0.0105(13)~fm$^2$ presented in~\cite{dr}. We are going to use our new number in further analysis.

 It is instructive  to explain why the field shift constants  $F$ have different signs for different electron states. Orbitals $s_{1/2}$ and $p_{1/2}$ penetrate nucleus and are highly sensitive to the nuclear radius (remind the reader that the lower  component of the Dirac spinor of the relativistic  $p_{1/2}$ orbital has angular quantum numbers of $s_{1/2}$ orbital). An increase of the nuclear radius  leads to decrease of the attraction  to the nucleus, therefore energies    $s_{1/2}$ and $p_{1/2}$  move up and constant $F$ is positive. Higher orbitals $p_{3/2}$,  $d$ and $f$ do not penetrate nucleus, so the direct term $\hat F$ in Eq.  (\ref{e:F}) is negligible. The effect comes from the correction to the electron core potential  $\delta V_{\rm core}$ which is dominated by the Coulomb field of $s_{1/2}$ electrons. Increase of the nuclear radius makes attraction to the nucleus  weaker, increases the radii of the $s_{1/2}$ orbitals and makes negative correction 
$\delta V_{\rm core}$   to the core electron Coulomb potential. This is why $F$ for   $p_{3/2}$,  $d$ and $f$ electrons is negative. We may also explain this sign from another end. Adding valence $p_{3/2}$,  $d$ or  $f$ electron increases positive Coulomb energy of the electron repulsion. As a result, the $s_{1/2}$ electron energies  and distances from the nucleus increase and their sensitivity to the change of the nuclear radius decreases. Thus, the effect of the higher wave valence  electron is negative.
% 10\% or better.
%Such accuracy is sufficiently good for the purpose of current work.

\begin{table} 
  \caption{\label{t:F}Field shift constant $F$ for the ground and some excited states of Th~IV, Th~III, Th~II, and Th~I.}
\begin{ruledtabular}
\begin{tabular}   {lll rrr}
\multicolumn{1}{c}{Atom}&
\multicolumn{2}{c}{State}&
\multicolumn{1}{c}{Expt. energy}&
\multicolumn{2}{c}{$F$ (GHz/fm$^2$)}\\
\multicolumn{1}{c}{or ion}&&&
\multicolumn{1}{c}{ (cm$^{-1}$)~\cite{Th-exp}}&
\multicolumn{1}{c}{Present}&
\multicolumn{1}{c}{Ref.\cite{dr}}\\
\hline
Th~IV  & $5f$       & $^2$F$^o_{5/2}$ &     0 & -55.0  &  \\
       & $5f$       & $^2$F$^o_{7/2}$ &  4325 & -53.0  &  \\
       & $6d$       & $^2$D$_{3/2}$   &  9193 & -23.3  &  \\
       & $6d$       & $^2$D$_{5/2}$   & 14586 & -20.5  &  \\
       & $7s$       & $^2$S$_{1/2}$   & 23130 &  92.1  &  \\
       & $7p$       & $^2$P$^o_{1/2}$ & 60239 &   2.7  &  \\
       & $7p$       & $^2$P$^o_{3/2}$ & 73055 &  -5.3  &  \\
       
Th~III & $5f6d$     & $^3$H$^o_4$     &     0 & -68.0 & -68.7 \\
       & $6d^2$     & $^3$F$_2$       &    63 & -39.9 & -36.6 \\
       & $5f^2$     & $^3$H$_4$       & 15148 & -83.3 & -89.5 \\
       & $5f6d$     & $^1$P$^o_1$     & 20711 & -62.2 & -63.6 \\
       & $5f^2$     & $^3$F$_4$       & 21784 & -86.5 & -85.5 \\
       & $5f^2$     & $^3$P$_0$       & 29300 & -82.2 & -84.1 \\

Th~II  & $6d^27s$   & $^2$D$_{3/2}$   &     0 &  49.6 &  54.6 \\
       & $5f6d^2$   & $*_{3/2}$       &       & -65.0 &       \\
       & $5f6d7s$   & $*_{3/2}$       & 15145 & -45.8 &       \\
       & $5f6d7s$   & $*_{3/2}$       & 15711 & -36.9 &       \\
       & $5f6d^2$   & $*_{3/2}$       & 17122 & -29.1 & -31.6 \\

       & $5f6d7s$   & $^2$F$^o_{5/2}$ & 12472 & -18.3 &  \\
       & $5f6d7s$   & $*_{5/2}$       &       & -36.3 &  \\
       & $5f6d7s$   & $^4$D$^o_{5/2}$ & 14545 & -63.9 &  \\
       & $5f6d7s$   & $*_{5/2}$       & 16033 & -46.8 &  \\

Th~I   & $6d^27s^2$ & $^3$F$_2$       &     0 &  58.6 &       \\

\end{tabular}			
\end{ruledtabular}
\end{table}

Using the  field shift constants for the ground states of each ion from Table~\ref{t:F} (we use our numbers for consistency), the value  
$\delta \langle r^2 \rangle^{229m,229}=0.0112(13)~{\rm fm}^2$ (see above) and formula similar to (\ref{e:ISN}) we obtain the differences between nuclear frequencies in different thorium ions. The results are presented in Table~\ref{t:dw}. 
We see that the difference is huge. It exceeds the projected relative uncertainty of the nuclear clocks by many orders of magnitude.
It is worth noting that the shift does not contribute to the uncertainty budget. It only means that the frequency of the nuclear transition is different in different thorium systems.

\begin{table}
  \caption{\label{t:dw}Change of nuclear frequency $\omega_N$ between ions of $^{229}$Th.}
\begin{ruledtabular}
\begin{tabular}   {lcl crr}
\multicolumn{3}{c}{Ions}&
\multicolumn{2}{c}{$\Delta \omega_N$}&
\multicolumn{1}{c}{$\Delta \omega_N/\omega_N$}\\
&&&\multicolumn{1}{c}{(GHz)}&
\multicolumn{1}{c}{(eV)}& \\
\hline
Th I  &$-$& Th II   &  0.10  & $4.3 \times 10^{-7}$ &   $5.0 \times 10^{-8}$ \\
Th II  &$-$& Th III & 1.3     &  $5.6 \times 10^{-6}$ &    $6.4 \times 10^{-7}$ \\
Th III &$-$& Th IV & -0.15  &   $-6.5 \times 10^{-7}$ &   $-7.3 \times 10^{-8}$ \\
Th II &$-$& Th IV & 1.2  &   $ 5.2\times 10^{-6}$ &   $ 5.9 \times 10^{-7}$ \\
\end{tabular}			
\end{ruledtabular}
\end{table}

It is interesting to determine the nuclear frequency difference between neutral (or nearly neutral) $^{229}$Th and bare $^{229}$Th nucleus.
This difference is strongly dominated by contributions from $1s$ electrons. Using the RPA calculation (\ref{e:RPA}) we get $F(1s) = 8.23 \times 10^8$ MHz/fm$^2$. The total energy shift caused be two $1s$ electrons is $1.73 \times 10^7$ MHz; the total shift from all core electrons is $2.07 \times 10^7$ MHz=8.57 $\cdot 10^{-2}$ eV,
which is $\sim$ 1\% of the nuclear frequency.

Electronic correction to the nuclear frequency comes also from magnetic interaction between electrons and nucleus. The first order gives ordinary magnetic hyperfine splitting of the transition frequencies. The magnetic shift  is given by the second-order magnetic dipole hyperfine correction to the energy
\begin{equation}\label{e:hfs}
%\delta E_g^{\rm hfs} = \sum_n \frac{\langle g | \hat H^{\rm hfs}|n\rangle^2}{E_g-E_n} \Delta\left(\frac{\mu}{I}\right).
\delta E_g^{\rm hfs} = \sum_n \frac{\langle g | \hat H^{\rm hfs}|n\rangle^2}{E_g-E_n}
\end{equation}
Here index $g$ stands for the ground electronic state, $\hat H^{\rm hfs}$ is the magnetic dipole hyperfine structure operator. Values  of $\delta E_g^{\rm hfs}$ are different for the ground  and isomeric nuclear states since their magnetic moments and spins are different. %The shift  $\Delta\left(\frac{\mu}{I}\right)$ is the difference between the ratios of nuclear magnetic moment $\mu$ to nuclear spin $I$ for nuclear ground and isomeric states.
Magnetic moment values  can be found in Ref.~\cite{Thm}.  In addition, there is the second order contribution from the mixing  of the ground and isomeric  nuclear states by the magnetic filed of electrons. Preliminary estimations show that the second order magnetic shift is significantly  smaller than the electronic shift considered in the present work. More detailed analysis might be a subject of further work.

\acknowledgments

This work was supported by the Australian Research Council Grants No. DP230101058 and DP200100150.

\end{document}